# Coherent radiation of microbunched beam in stack of two plates

**Hayk Gevorgyan**[1,2,*], **Lekdar Gevorgyan**[1] and **Karo Ispiryan**[1]

[1] A.Alikhanyan National Laboratory (Yerevan Physics Institute), Br. Alikhanyans 2, Yerevan, 0036, Armenia

[2] Yerevan State University (YSU), 1 Alex Manoogian, Yerevan, 0025, Armenia

E-mail: haykgev95@gmail.com

**Abstract**. The present article is devoted to the radiation from an electron bunch with modulated density passes through the stack consisting of two plates with different thicknesses and electrodynamic properties. The new elegant expression for the frequency-angular distribution of transition radiation is obtained. Using the existence of resonant frequency at which the longitudinal form-factor of bunch not suppresses radiation coherence and choosing parameters for the stack of plates, one can also avoid suppression of the radiation coherence by transverse form-factor of bunch. The radiation from a bunch with modulated density in the process SASE FEL can be partially coherent at a resonant frequency. Then the intense sub monochromatic beam of X-ray photons is formed. On the other hand one can define an important parameter of density modulation depth of bunch which is unknown to this day.

## 1. Introduction

The phenomenon of short wavelength coherent radiation of asymmetric bunch is discovered in the work [1]. The radiation intensity is proportional to the second power of particle number of bunch. This phenomenon is universal and does not depend on the kind of radiation. The effect of coherence of diffraction radiation was observed in the Japanese experiment [2]. By radiation spectrum one can determine the real distribution of electrons in the bunch. The use of truncated electron bunch will significantly increase the free electron laser (FEL) efficiency [3] and will fulfill the coherence of diffractive [4] and Smith-Parcell [5] radiations.

In the work [6] is researched the radiation problem of the bunch which density is modulated by laser beat waves (LBW). It is shown that the longitudinal form-factor of the bunch modulated at the resonance frequency $\omega_r$ does not suppress the coherence effect. The suppression of coherence may occur due to the transverse form-factor.

The electron bunch which interacts with LBW coherently radiates in a spiral undulator in the sub-millimeter range [7]. In the work [8] was offered the use of partially coherent radiation at a resonant frequency for definition of the modulation depth of LCLS (linac coherent light source) bunch.

The results of article [9] show that one can use X-ray crystalline undulator radiation for study of the microbunching process in XFEL and for the production of monochromatic intense beam. A review on the coherent X-ray radiation of various types produced by the microbunched beam in amorphous and crystalline radiators [10] is given.

This work is devoted to the coherent X-ray radiation from a bunch of microbunched electrons produced in the stack of two plates at a given parameters.

## 2. Coherent radiation of electron bunch

In the work [1] is received the formula for frequency-angular average distribution of the number of photons of any nature and emitted with freeform electron bunch:

$$N_{tot}(\omega,\vartheta) = <N_{ph}(\omega,\vartheta)> = N_{sp}(\omega,\vartheta)\left(N_b^2 F + N_b(1-F)\right), \quad (1)$$

where $N_{sp}(\omega,\vartheta)$ is the frequency-angular distribution of photons number of single electron radiation, $N_b$ is the number of electrons in the bunch, and the bunch form-factor $F$ is determined with longitudinal $F_Z(\omega)$ and transverse $F_R(\omega,\vartheta)$ form-factors by the formulas:

$$F = F_Z(\omega) \cdot F_R(\omega,\vartheta), \quad (2)$$

$$F_Z(\omega) = \left|<exp[-ik_\parallel Z]>\right|^2 = \left|\int f(Z)\,exp[-ik_\parallel Z]\,dZ\right|^2,$$

$$F_R(\omega,\vartheta) = |<exp[-ik_\perp R]>|^2 = |\int f(R)\,exp[-ik_\perp R]\,dR|^2.$$

Here the averaging is done by the functions $f(Z)$ and $f(R)$ of the electron bunch density distribution in longitudinal $Z$ and transverse $R$ directions. If $f(R)$ is Gaussian distribution function with dispersion $\sigma_R^2$, then for $F_R$ we have:

$$F_R(\lambda,\vartheta) = \exp\left[-\left(\frac{2\pi\sigma_R\vartheta}{\lambda}\right)^2\right], \quad (3)$$

where $\lambda$ is the radiation wavelength.

When $F_Z$ is asymmetrically-Gaussian with dispersions $\sigma_Z^2$ and $(p\sigma_Z)^2$, then is received following formula:

$$f(Z) = \frac{2}{(1+p)\sqrt{2\pi}\sigma_Z}\left(\exp\left[-\frac{1}{2}\cdot\left(\frac{Z}{\sigma_Z}\right)^2\right]\theta(-Z) + \exp\left[-\frac{1}{2}\cdot\left(\frac{Z}{p\sigma_Z}\right)^2\right]\theta(Z)\right). \quad (4)$$

where the parameter $p$ ($0 \leq p \leq 1$) determines the degree of asymmetry, $\theta(Z)$ is the Heaviside unit step function:

$$\theta(Z) = \begin{cases} 1, & for \;\; Z \geq 0 \\ 0, & for \;\; Z < 0 \end{cases}. \quad (5)$$

For the longitudinal form-factor is obtained the following expressions:

$$F_Z(t) = \frac{1}{(1+p)^2}\left(\left(e^{-t^2} + pe^{-p^2 t^2}\right)^2 + \frac{4}{\pi}\left(W(t) - pW(pt)\right)^2\right), \quad (6)$$

$$W(t) = \int_0^t exp[x^2 - t^2]dx, \quad t = \frac{\sqrt{2}\pi\sigma_Z}{(1+p)\lambda}.$$

For a symmetrical distribution ($p = 1$) we have:

$$F_Z(t) = e^{-2t^2} = e^{-\left(\frac{2\pi\sigma_Z}{\lambda}\right)^2} = e^{-\eta}. \quad (7)$$

For the short wavelengths ($\lambda \ll \pi\sigma_Z$) the form-factor $F_Z(\lambda)$ decreases exponentially although $F_R(\lambda,\vartheta) = 1$. When the distribution is asymmetric function ($0 \leq p < 1$), the exponentially decreasing is replaced by the power-law. When the derivative $f^{(n)}(Z)$ is discontinuous at point $Z_0$ and $f(Z_0)$ is the maximum value of function $f(Z)$, then the asymptotic estimate is:

$$F_Z(\eta) \sim \eta^{-2(n+1)}. \quad (8)$$

In the case, when the bunch density distribution $f(Z)$ has a form (4), the longitudinal form-factor is decreased as power-law $\eta^{-6}$ ($n = 2$). If the electron bunch density distribution is discontinuous function ($n = 0$), then the form-factor decreases as power-law $\eta^{-2}$.

### 3. Generated X-ray radiation in stack of two plates

In the work: Gevorgian L.A. "Garibyan Radiation as a Source of X-ray Photons" (to be published) is considered the problem of radiation when an electron passing through the stack consisting of two different plates with the thicknesses $a$, $b$ and the plasma frequencies of media $\omega_1$, $\omega_2$. It was defined the average plasma frequency $\omega_p$ of stack and the parameter $\Delta$, characterizing the difference between electrodynamical properties in media as follows:

$$\omega_p^2 = \frac{\omega_1^2 + \omega_2^2}{2}, \qquad 0 < \Delta = \frac{\omega_1^2 - \omega_2^2}{\omega_1^2 + \omega_2^2} \ll 1. \tag{9}$$

For the frequency-angular distribution of spontaneously radiated photons as the function of dimensionless frequency $x = \omega/(\omega_p \gamma)$ and $u = \vartheta \gamma$ ($\gamma$ is the Lorentz-factor) is received:

$$N_e^{sp}(x, u) \equiv \frac{d^2 N_e^{sp}}{dx\, du^2} = \frac{32 \Delta^2 np}{137\pi} \frac{u^2 x^2 \sin^2\left(\frac{\pi}{2} z(x, u)\right)}{((x^2(u^2 + 1) + 1)^2 - \Delta^2)^2} \delta(u^2 - \varphi(x)), \tag{10}$$

$$z(x, u) = \frac{a}{\lambda_p \gamma} \frac{x^2(1 + u^2) + 1 - \Delta}{x}, \qquad \varphi(x) = \frac{2p}{x} - 1 - \frac{1}{x^2},$$

$$p = \frac{\gamma}{\gamma_{th}}, \qquad \gamma_{th} = \frac{l}{\lambda_p}, \qquad n = \frac{L}{l}, \qquad l = a + b,$$

where $n$ is the number of pairs of plates, $L$ is the stack length, $\lambda_p$ is the average plasma wavelength of stack, $\gamma_{th}$ is the energy threshold for formation of transition radiation, and $\delta(u^2 - \varphi(x))$ is the Dirac's delta function.

The frequency distribution of number of radiated photons, with accuracy to small $\Delta^2$, can be represented:

$$\frac{dN_e^{sp}}{dx} = \frac{32 \Delta^2 pn}{137\pi} \frac{(x - x_1)(x_2 - x)}{(2px)^4} \sin^2\left(\frac{\pi}{2} z(x)\right), \tag{11}$$

$$z(x) = 1 - \xi + \frac{\Delta}{2px}, \qquad \xi = \frac{a - b}{l}.$$

Note that the distribution is significantly different from zero at $p \gg 1$, then the frequency interval of radiation is:

$$\frac{1}{2p} = x_1 \le x \le x_2 = 2p. \tag{12}$$

Considering the above, the radiation in the stack is constructively for all values of $x$. The radiation is formed when $\gamma > \gamma_{th}$. Then we have

$$N_e^{sp}(x) \equiv \frac{dN_e^{sp}}{dx} = \frac{32 \Delta^2 pn}{137\pi} \frac{(x - x_1)(x_2 - x)}{(2px)^4}. \tag{13}$$

The maximum of distribution is at the frequency $x_m = 2/(3p)$. Herewith choice of values for the parameters $a$ and $b$ follows from the condition $y(x_m) = 1$ or $\xi = 3\Delta/4$, and therefore:

$$l = \frac{3}{2} \frac{\lambda_p^2}{\lambda_m}, \qquad a = \frac{l}{2}\left(1 - \frac{3}{4}\Delta\right), \qquad b = \frac{l}{2}\left(1 + \frac{3}{4}\Delta\right). \tag{14}$$

If the parameters are chosen according to expression (6), then the radiation in stack will be constructively at the frequency $x_m = \lambda_p/(\lambda_m \gamma)$ and occurs at angle $\vartheta_m = 1/\sqrt{3} x_m \gamma$.

The frequency distribution is represented conveniently as a function from the variable $y = x/x_m$. On condition $\gamma \gg \gamma_{th}$ it follows that $x_1 \ll 1$ or $x_2 = 1/x_1 \gg 1$ and the frequency distribution has the following form:

$$N_e^{sp}(y) = \frac{9\Delta^2 n}{4\cdot 137\pi} g(y), \qquad g(y) = \frac{1}{y^3}\left(4 - \frac{3}{y}\right). \tag{15}$$

Consequently, $g(0.75) = 0$, $g(1) = 1$ and the distribution function $g(y)$ is presented in Figure 1.

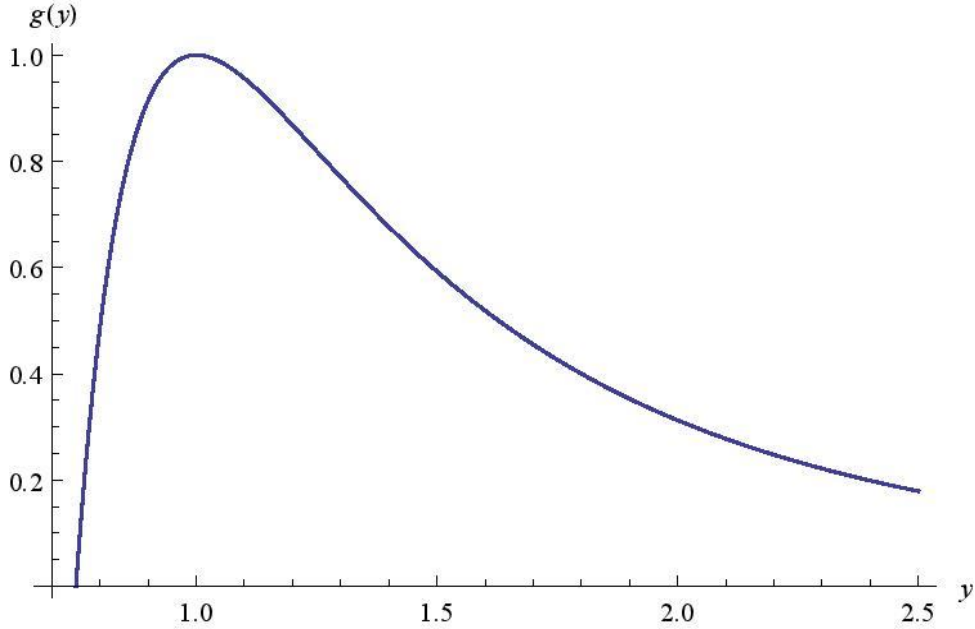

**Figure 1.** The frequency distribution of radiation generated in the stack of two plates.

The total number of radiation photons is equal to

$$N_{tot}^{sp} = N_b N_e^{sp} = \frac{8}{3} N_b \frac{\Delta^2 n}{137\pi}. \tag{16}$$

### 4. Coherent radiation of beam with modulated density at the resonant frequency

If we assume that the bunch density is modulated by law $2b_1 \cos(2\pi Z/\lambda_r)$, where $2b_1$ is the modulation depth, then the frequency distribution of coherent radiation which is formed in the stack of the plates by the bunch of $N_b$ electrons has the following form:

$$N_{coh}(x) = b_1^2 N_b^2 N_e^{sp}(x)\, exp[-A(x - x_r)^2]\, exp[-B(x_r - x_1)(x_2 - x_r)], \tag{17}$$

$$A = \left(\frac{2\pi \sigma_Z \gamma}{\lambda_p}\right)^2, \qquad B = \left(\frac{2\pi \sigma_R}{\lambda_p}\right)^2.$$

Here is considering that the value of A is much more B ($A \gg B \gg 1$).

If choose the resonance frequency $x_r$ taking close to $x_1$, i.e. $x_r = x_1(1 + 1/B)$, then the transverse form-factor is equal to $1/e$. For this it's necessary have the stack of plates with the parameters

$$l = \frac{2\lambda_p^2}{\lambda_r}, \qquad a = \frac{l}{2}(1 - \Delta), \qquad b = \frac{l}{2}(1 + \Delta). \tag{18}$$

Then we have:

$$y(x_r) = 1, \quad 2px_r = 1, \quad (x_r - x_1)(x_2 - x_r) = \frac{1}{B}. \tag{19}$$

And follows the value for number of radiated photons at resonance frequency:

$$N_e^{sp}(x_r) = \frac{16\Delta^2 n}{137\pi B x_r} = K_{sp} \frac{n\Delta^2 \lambda_p \lambda_r}{\sigma_R^2}, \tag{20}$$

$$K_{sp} = \frac{4}{137\pi^3} \approx 9.4 \cdot 10^{-4}, \quad \vartheta_r = \frac{\lambda_r}{\pi \sigma_R},$$

where $\vartheta_r$ is the radiation angle.

The coherent radiation distribution is defined by the longitudinal form-factor. For the frequency distribution of coherent radiation we have

$$N_{coh}(x) = K_{coh} n \lambda_p \lambda_r \gamma \frac{L}{\sigma_Z} \left(\frac{\Delta b_1 N_b}{\sigma_R}\right)^2 F_{coh}(x), \tag{21}$$

$$F_{coh}(x) = exp[-A(x - x_r)^2], \quad K_{coh} = 3.464 \cdot 10^{-4}.$$

The total number of coherent radiated photons is the following:

$$N_{tot}^{coh} = k_{coh} \frac{L}{\sigma_Z} \left(\frac{\Delta \lambda_r b_1 N_b}{\sigma_R}\right)^2 \tag{22}$$

with the line width of $2/\sqrt{A}$ and the radiation angle of $\lambda_r/(2\pi\sigma_R)$.

## 5. Parameters corresponding to LCLS SASE FEL

Let's calculate the number of generated X-ray photons formed in stack of plates with selected parameters $a$ and $b$ which is located at the end of SASE FEL.

The parameters of SASE (self-amplified spontaneous emission) FEL are:
$$\lambda_r = 1.5 \cdot 10^{-8} \text{ cm } (\hbar\omega_r = 8.3 \text{ KeV}), \quad b_1.$$

The beam parameters of microbunched beam of LCLS are:
$$E = 13.6 \text{ GeV}, \quad N_b = 1.56 \cdot 10^9, \quad \sigma_Z = 9 \cdot 10^{-4} \text{ cm}, \quad \sigma_R = 6.12 \cdot 10^{-4} \text{ cm}, \quad b_1.$$

The stack parameters are:
$$\Delta = 2/9 \quad (\hbar\omega_1 = 30 \text{ eV}, \hbar\omega_2 = 24 \text{ eV}, \hbar\omega_p = 27 \text{ eV}), \quad \lambda_p = 4.6 \cdot 10^{-6} \text{ cm},$$

the space period for the coherent radiation is equal to $l = 28$ μm ($a = 11$ μm, $b = 17$ μm).

The thickness of stack is $L = 2.8$ cm for the following condition: $L \lesssim L_{abs}$, where $L_{abs}$ is the length of absorption.

The photon number of the coherent radiation is equal to $4.8 \cdot 10^9 b_1^2$, the line width is $6.1 \cdot 10^{-8}$, the radiation angle is $3.9 \cdot 10^{-6}$.

Then we have

$$\frac{N_{coh}}{N_{sp}} = 1.17 \cdot 10^8 b_1^2 \gg 1, \quad \text{if } b_1 \gg 10^{-4}.$$

These results show that one can use for study of the microbunching process in XFEL and for the production of monochromatic intense photon beams.

## 6. Conclusion

For the frequency distribution of the radiated photon number from a single electron transition radiation the simple formula is received due to constructive interference.

The spectral-angular distribution of microbunched beam is presented of a product of the corresponding quantity for a single electron and a bunch form-factor.

The longitudinal form-factor of the modulated beam not suppresses radiation coherence at the beam resonance frequency. The applications of these new results for some experimental possibilities are discussed. One can also avoid to suppression of the radiation coherence at resonant frequency by transverse form-factor. If the electron beam, modulated in the process SASE FEL passes through the stack of plates then at a resonant frequency the coherent submonochromatic radiation is generated. It is shown that one can use the results for study of the microbunching process in XFEL and for the production of monochromatic intense photon beams.